\documentclass[9pt,twocolumn,twoside]{opticajnl}
\journal{opticajournal} 

\setboolean{shortarticle}{true}

\usepackage{lineno}

\title{Continuous wave driving elucidates the desynchronisation dynamics of ultrashort dissipative Raman solitons generated in dispersive Kerr resonators}

\author[1,2,*]{Zongda Li}
\author[1,2]{Yiqing Xu}
\author[1,2]{Stéphane Coen}
\author[1,2]{Stuart G. Murdoch}
\author[1,2]{Miro Erkintalo}

\affil[1]{Department of Physics, University of Auckland, Auckland, New Zealand}
\affil[2]{The Dodd-Walls Centre for Photonic and Quantum Technologies, Dunedin,
New Zealand}
\affil[*]{zongda.li@auckland.ac.nz}

\begin{abstract}
Phase-coherent pulsed driving of passive optical fiber resonators enable the generation of ultrashort dissipative Raman solitons with durations well below 100~fs. The existence and characteristics of such solitons critically depends on the desynchronisation between the pulsed driving source and the resonator roundtrip time, yet the full mechanism through which these dependencies arise remains unclear. Here, we numerically demonstrate that Raman solitons can exist even under conditions of continuous wave driving, and by numerically examining the existence and characteristics of Raman solitons under such conditions, we elucidate the role of desynchronisation in pulse-driven systems. In addition to providing new insights on the existence and characteristics of ultrashort Raman solitons, our analysis yields a qualitative explanation for the range of desynchronisations over which the solitons can exist.

\end{abstract}

\setboolean{displaycopyright}{false} 

\begin{document}

\maketitle

\noindent Dispersive Kerr resonators driven with coherent laser light host a variety of temporal localized structures. Arguably the most well-known of these is the bright Kerr cavity soliton (CS)~\cite{leo_temporal_2010, gaeta_photonic-chip-based_2019}, but other examples include dark solitons~\cite{xue_mode-locked_2015}, platicons~\cite{jang_dynamics_2016}, Dirac solitons~\cite{wang_dirac_2020}, parametrically driven cavity solitons~\cite{englebert_parametrically_2021}, and Stokes solitons~\cite{yang_stokes_2017}. These structures have attracted substantial interest over the past decade, not only due to their rich nonlinear dynamics, but also because of their practical utility. Indeed, in the spectral domain, temporal localized structures correspond to coherent (Kerr) frequency combs~\cite{coen_modeling_2013,herr_temporal_2014}, whose many applications range from telecommunications to spectroscopy~\cite{suh_microresonator_2016, marin-palomo_microresonator-based_2017, trocha_ultrafast_2018, suh_searching_2019}.

A notable structure that has recently been discovered in driven Kerr resonators is the ultrashort dissipative Raman soliton ~\cite{xu_frequency_2021,li_ultrashort_2023}. Such Raman solitons have been experimentally shown to arise via a complex interplay between dispersion, Kerr nonlinearity, and stimulated Raman scattering, when a resonator with a broad Raman gain spectrum is driven under appropriate conditions with phase-coherent optical pulses. Compared to conventional CSs in similar systems, Raman solitons can exhibit significantly shorter durations, with solitons as short as 50~fs realized in meter-scale resonators made from standard telecommunications optical fiber~\cite{li_ultrashort_2023}. Accordingly, Raman solitons provide an interesting avenue to generate broadband frequency combs with sub-GHz and GHz repetition rates, which could be useful for applications such as high resolution spectroscopy. 

The desynchronisation between the periodicity of the driving pulse train and the resonator roundtrip time is a key control parameter that governs the existence and characteristics of Raman solitons. In ref.~\cite{li_ultrashort_2023}, the impact of desynchronisation on the soliton characteristics was qualitatively explained through the phase- and group-velocity-matching conditions that must be satisfied, but this analysis does not reveal insights on the range of desynchronisations over which the solitons can exist. In this Letter, we fill this gap through extensive numerical modelling. Specifically, we use numerical simulations to show that Raman solitons can exist even under conditions of continuous wave (CW) driving, and that the soliton characteristics -- including group-velocity -- depend upon the intracavity background power that the soliton experiences. Under conditions of pulsed driving, the soliton is attracted to, and trapped at, an instantaneous background power level associated with a group-velocity that exactly cancels the desynchronisation; adjusting the desynchronisation changes the trapping power, hence modifying the soliton characteristics. In addition to elucidating the role of pump-cavity desynchronisation in the generation of ultrashort Raman solitons, our analysis explains why the solitons can only exist over a finite range of desynchronisations.

We begin by recounting the key physics of ultrashort Raman solitons. To this end, we perform numerical simulations using the following generalized Lugiato-Lefever equation~\cite{li_ultrashort_2023}, which describes the evolution of the slowly-varying electric field envelope, $E(t,\tau)$, within a dispersive Kerr resonator:
\begin{align}
	t_\mathrm{R}\frac{\partial E}{\partial t} =& \left[-\alpha-i\delta_0-\Delta t \frac{\partial}{\partial \tau} + i L \hat{D}\left(i\frac{\partial}{\partial\tau}\right) \right]E + \sqrt{\theta P_\mathrm{P}} S(\tau) \nonumber \\
	& +i\gamma L \left[(1-f_\mathrm{R})|E|^2 + f_\mathrm{R}h_\mathrm{R}(\tau)\ast|E|^2 \right]E. \label{LLE}
\end{align}
Here, $t$ is the ``slow'' time variable that describes the evolution of the field envelope $E(t,\tau)$ over multiple roundtrips, whilst $\tau$ is a corresponding ``fast'' time that describes the envelope's profile over a single roundtrip. $\alpha = \pi/\mathcal{F}$ describes the resonator losses with $\mathcal{F}$ the resonator finesse, $\delta_0$ is the linear phase detuning between the carrier frequency of the driving pulse train and a cavity resonance, $\Delta t = t_\mathrm{R}-t_\mathrm{P}$ is the per-roundtrip desynchronisation between the period of the driving pulse train ($t_\mathrm{P}$) and the resonator roundtrip time ($t_\mathrm{R}$)~\cite{coen_convection_1999}, $L$ is the roundtrip length of the resonator, $\hat{D}(i\partial/\partial\tau) = \sum_{k>2} \beta_k/k! (i\partial/\partial\tau)^k$ is the linear dispersion operator with $\beta_k$ the Taylor series expansion coefficients of the resonator propagation constant about the carrier frequency of the driving pulse train, $\theta$ is the coupling coefficient with which the driving pulses with peak power $P_\mathrm{P}$ and temporal profile $S(\tau)$ are injected into the resonator [$\text{max}\{S(\tau)\}=1$], $\gamma$ is the Kerr nonlinearity coefficient, $f_\mathrm{R}$ is the Raman fraction, and $h_\mathrm{R}(\tau)$ is the time-domain Raman response function~\cite{hollenbeck_multiple-vibrational-mode_2002}. 

\begin{figure}[h!]
\centering\includegraphics[scale = 1]{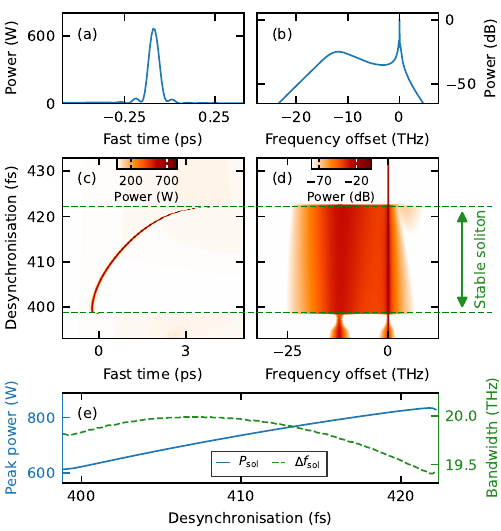}
\caption{Numerical simulation results, illustrating the key characteristics of Raman solitons. All simulations use $\mathcal{F}=150$, $\theta = 0.01$, $\gamma = 1.8$~W$^{-1}$km$^{-1}$, $f_\mathrm{R} = 0.09$, $P_\mathrm{p} = 100$~W, $\delta_0 = 0$, $\beta_2 = 7.15$~ps$^2$km$^{-1}$, $\beta_3 = 0.14$~ps$^3$km$^{-1}$, and $\beta_k = 0$ for $k> 3$. (a, b) Blue curves show steady state (a) temporal and (b) spectral intensity profiles for a desynchronization $\Delta t = 404$~fs. (c, d) steady-state (c) temporal and (d) spectral intensity profiles as a function of desynchronization. The dashed green lines demarcate the desynchronisation range over which stable Raman solitons exist. (e) The peak power of the soliton $P_\mathrm{sol}$ (solid blue, left axis) and spectral bandwidth of the soliton $\Delta f_\mathrm{sol}$ (dashed green, defined as $-20$~dB below the peak of soliton spectral components, right axis) as a function of desynchronisation, extracted from (c) and (d), respectively.}\label{fig1}
\end{figure}

Figures~\ref{fig1}(a) and (b) show illustrative temporal and spectral intensity profiles of a steady-state Raman soliton, respectively, obtained by numerically integrating~\eqref{LLE} with experimental parameters (extracted from ref.~\cite{li_ultrashort_2023}) corresponding to a 2.65-m-long resonator made from dispersion-shifted optical fibre (Corning MetroCor) driven with Gaussian pulses with 4-ps-duration [for other parameters, see caption of Fig.~\ref{fig1}]. The steady-state profiles reveal a 82-fs-duration soliton with a broad spectrum that bridges the driving frequency and the peak of the Raman gain spectrum. 

To illustrate the solitons' dependence on the desynchronisation $\Delta t$, the pseudocolor plots in Figs.~\ref{fig1}(c) and (d) show numerically simulated steady-state intracavity temporal and spectral intensity profiles for a range of desynchronisations, respectively. As can be seen, the Raman solitons only exist over a finite range of desynchronisations, and their spectral bandwidth, peak intensity, and temporal delay $\tau_\mathrm{s}$ with respect to the peak of the driving pulse (at $\tau = 0$) all change with desynchronisation [see also Fig.~\ref{fig1}(e)]. In what follows, we will show that these changes (as well as the finite desynchronisation range over which the solitons exist) can be understood by considering the system dynamics under conditions of CW driving.

In contrast to pulse-driven systems, stable Raman solitons do not form spontaneously (i.e., starting from an empty cavity) with CW driving. This is because spontaneous soliton emergence requires the intracavity background prior to soliton formation to be above the Raman lasing threshold~\cite{agrawal-nonlinear-fibre-optic}. In pulse-driven systems, only the peak of the intracavity background can be configured to be above this threshold, thus yielding isolated and stable solitons, but this is not possible in CW-driven systems. However, while the solitons' \emph{spontaneous generation} requires the intracavity background to be above the Raman threshold, their \emph{existence} does not. If externally seeded, Raman solitons can exist even at driving power levels that are globally below the Raman threshold; it is this hysteresis behaviour that permits the solitons to exist with CW driving.

\begin{figure}[h!]
\centering\includegraphics[scale = 1]{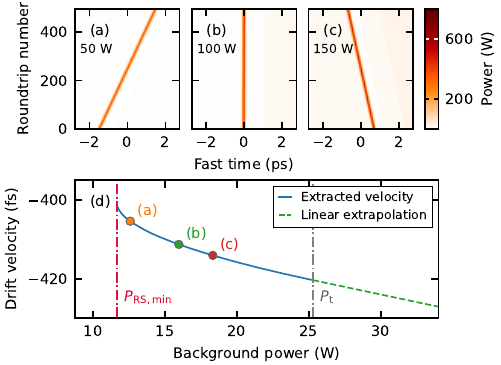}
\caption{(a)--(c) Numerically simulated roundtrip-to-roundtrip temporal evolution of a Raman soliton driven by a CW beam with a power of (a) 50~W, (b) 100~W, and (c) 150~W. (d) Raman soliton temporal drift velocity in the natural reference frame of the resonator as a function of background power. The minimum background power $P_\mathrm{RS,min}$ required to sustain solitons, and the CW Raman gain threshold $P_\mathrm{t}$, are indicated by the dashed-dotted red and grey lines, respectively. The dashed green curve shows a linear extrapolation of the drift velocity for background powers $P_\mathrm{B} > P_\mathrm{t}$. Solid circles in (b) highlight the background powers associated with panels (a)--(c).}\label{fig2}
\end{figure}

To illustrate the salient dynamics of Raman solitons with CW driving, we use \eqref{LLE} with $S(\tau) = 1$ and an initial condition close to a Raman soliton extracted from pulse-driven simulations. We find that, in steady-state, the solitons' characteristics -- including group-velocity -- depend upon the CW background power level, defined as the CW steady-state of the system in the absence of a soliton (or equivalently the CW power that surrounds the soliton when present). This is illustrated in Fig.~\ref{fig2}, which shows typical simulation results obtained with different CW driving power levels ($P_\mathrm{P}$) and with all other parameters constant (and as in Fig.~\ref{fig1}). 

\begin{figure}[h!]
\centering\includegraphics[scale = 1]{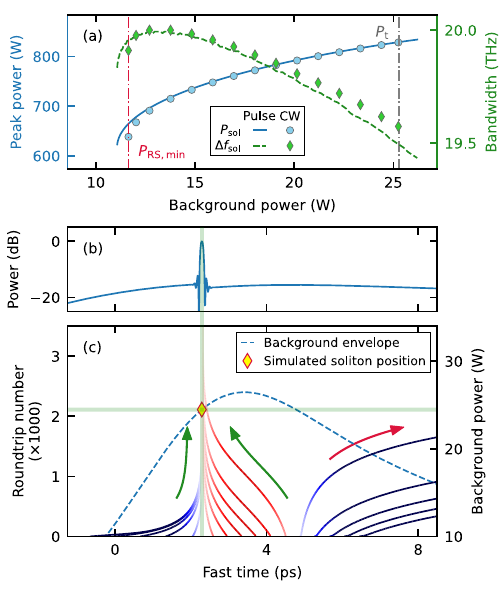}
\caption{(a) Solid blue and dashed green curves show Raman soliton peak power (left axis) and bandwidth (right axis) as a function of background power, respectively, extracted from simulations in Fig.~\ref{fig1}. The solid circles and diamond markers show corresponding results from CW simulations. (b) Steady-state temporal intensity profile of a pulse-driven Raman soliton with $\Delta t = 420$~fs. (c) Simulated response of the Raman soliton in (b) when temporally displaced by various amounts. The solid lines show the roundtrip-to-roundtrip evolution of the soliton's temporal position, with red (blue) indicating the soliton is sitting atop an instantaneous background power level above (below) the steady-state value. The dashed curve in (c) shows the steady-state background envelope prior to significant growth of Raman signal. The yellow diamond in (c) indicates the predicted critical position that cancels desynchronization, whilst the horizontal and vertical green lines indicate the instantaneous background power and the temporal position of the simulated steady-state soliton.}\label{fig3}
\end{figure}

Figures~\ref{fig2}(a)--(c) show the numerically simulated temporal intensity profiles of CW-driven Raman solitons over consecutive roundtrips for three different driving power  levels. We observe clearly a temporal drift with a rate per roundtrip, $V(P_\mathrm{B})=t_\mathrm{R}d\tau_\mathrm{s}/dt$,  that depends on the background power level $P_\mathrm{B}$. (To facilitate visualisation, the trajectories are displayed in a reference frame where the soliton in Fig.~\ref{fig2}(b) is stationary.) To gain more insights, we extract the drift velocity (in the resonator's natural reference frame) over a range of background power levels, and depict the results in Fig.~\ref{fig2}(d).  As can be seen, the curve is monotonically decreasing with $V(P_\mathrm{B})<0$ for all $P_\mathrm{B}$. Besides affecting the solitons' temporal drift (or group-velocity), we also find that the background power level influences soliton's peak power and spectral width [c.f. Fig.~\ref{fig3}(a)].

The CW-driving results described above suggest an explanation for the mechanism through which desynchronisation affects the existence and characteristics of Raman solitons in pulse-driven systems. We specifically hypothesize that, due to their ultrashort durations, the solitons respond (at least to first order) to the instantaneous background that they sit atop; in other words, each temporal position, $\tau$, along the inhomogeneous intracavity background (that is due to desynchronized pulsed driving) can be identified with a single background power level $P_\mathrm{B}(\tau)$ associated with a single set of soliton characteristics, which includes the temporal drift rate $V(P_\mathrm{b}(\tau))$. Thus, for the solitons to be stationary (as necessary for steady-state conditions), they must sit at a critical temporal position, $\tau_\mathrm{C}$, associated with an instantaneous background power where the drift velocity cancels the desynchronisation, i.e., $V(P_\mathrm{b}(\tau_\mathrm{C})) = -\Delta t$. Changing the desynchronisation is expected to change the solitons' steady-state temporal position -- as indeed observed in Fig.~\ref{fig1}(c) -- as well as their characteristics due to a change in the instantaneous background power they experience. 

Figure~\ref{fig3} shows results that corroborate the hypothesis describe above. Here, Fig.~\ref{fig3}(a) compares the soliton characteristics from Fig.~\ref{fig1}(e) plotted as a function of the instantaneous background power level associated with each desynchronisation with the same characteristics extracted from CW-driving simulations. The two sets of data are clearly in good agreement, with small discrepancies attributed to the gradient of the background profile in the pulse-driven scenario. 

Because the drift velocity $V(P_\mathrm{B})$ decreases monotonously with the background power $P_\mathrm{B}$ [c.f. Fig.~\ref{fig2}(d)], a typical single-peaked (unimodal) intracavity background profile will admit zero or two critical temporal positions that satisfy the condition $V(P_\mathrm{b}(\tau_\mathrm{C})) = -\Delta t$ (one on the leading and the other on the trailing edge of the pulse). However, it is straightforward to show that only the leading edge position is stable. Effectively, the critical temporal position $\tau_\mathrm{C}$ in the leading edge of the pulse acts as a robust trapping point, in analogy with the trapping of conventional CSs by phase or intensity modulated driving fields~\cite{jang_temporal_2015,hendry_spontaneous_2018}. This is demonstrated by the simulation results shown in Figs.~\ref{fig3}(b) and (c). Here we consider pulse-driven Raman solitons with parameters identical to those in Fig.~\ref{fig1}(a) and (b). We perform multiple simulations, each of which start from an initial condition where the steady-state Raman soliton [shown in Fig.~\ref{fig3}(b)] is displaced by various amounts from the stationary position, and depict in Fig.~\ref{fig3}(c) the evolution of the soliton positions as solid lines. The solitons are attracted towards the leading edge critical position, but repelled away from the corresponding position in the trailing edge. Also shown as the blue curve in Fig.~\ref{fig3}(c) is the intracavity background intensity profile prior to soliton formation. For the parameters of the simulation, the critical background power required to cancel the desynchronisation is found from CW-driving simulations to be $P_\mathrm{B} = 24.8~\mathrm{W}$, which is reached at $\tau_\mathrm{C} = 2.3$~ps. This point [yellow diamond in Fig.~\ref{fig3}(c)] is found to be close to the actual stable soliton position observed in pulse-driven simulations. 

The results presented above demonstrate that desynchronisation affects the characteristics of Raman solitons primarily by changing the soliton's steady-state temporal position, which in turn changes the instantaneous background power that the solitons experience. This explanation also allow us to readily understand why Raman solitons only exist over a finite range of desynchronisations. The upper limit of the desynchronisation range arises because a given background intensity profile can only compensate for desynchronisations $\Delta t \leq \Delta t_\mathrm{max}= -\text{min}\{V(P_\mathrm{B}(\tau))\}$. Because the drift velocity decreases monotonously with the background power $P_\mathrm{B}$, the maximum desynchronisation is achieved when the soliton is trapped at the peak of the intracavity background envelope: $\Delta t_\mathrm{max} = -V(P_\mathrm{B, peak})$. In other words, increasing the desynchronisation shifts the critical temporal position towards the peak of the background envelope, beyond which no stable positions are available [c.f. Fig.~\ref{fig1}(c)]. The lower limit of the desynchronisation range ($\Delta t_\mathrm{min}$) arises because solitons can only exist at background power levels above a given threshold [e.g. $P_\mathrm{RS,min}$ in Fig.~\ref{fig2}(d)]. The solitons' drift velocity is maximised at that threshold (recall that $V(P_\mathrm{B})<0$), which corresponds to the minimum tolerable desynchronisation: $\Delta t_\mathrm{min} = -V(P_\mathrm{RS,min})$. 

\begin{figure}[h!]
\centering\includegraphics[scale = 1]{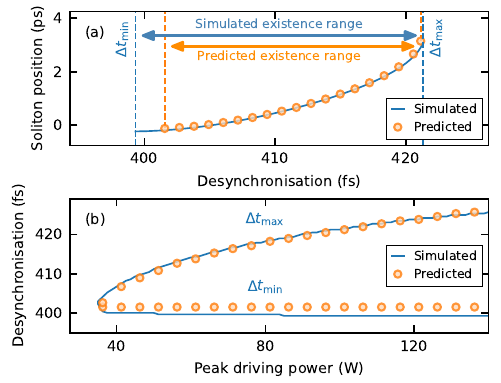}
\caption{(a) Solid blue curve shows the steady-state soliton position as a function of desynchronization extracted from Fig.~\ref{fig1}(c). The orange solid circles show corresponding positions obtained by solving the equation $V(P_\mathrm{B}(\tau_\mathrm{C})) = -\Delta t$. The dashed blue and orange lines highlight the soliton existence boundaries observed in pulse-driven simulations and predicted from CW analysis, respectively. (b) Blue solid curves show the minimum and maximum desynchronizations between which steady-state Raman solitons exist as a function of peak driving power, whilst the orange solid circles show the corresponding desynchronization ranges predicted from CW analysis.}\label{fig4}
\end{figure}

Figure~\ref{fig4} shows  results that support the above interpretation. The solid curve in Fig.~\ref{fig4}(a) shows the steady-state temporal position of the Raman solitons [extracted from data in Fig.~\ref{fig1}(c)] as a function of desynchronisation, whilst solid circles show the corresponding positions obtained by solving the equation $V(P_\mathrm{B}(\tau_\mathrm{C})) = -\Delta t$, with the curve $V(P_\mathrm{B})$ extracted from the CW-driving simulations in Fig.~\ref{fig2}(d). The two sets of data are in good agreement. The only notable discrepancy is the  lower limit of the existence range, which is under-estimated by the CW analysis; it appears that the gradient of the background allows solitons to exist for instantaneous power levels somewhat lower than the minimum threshold value observed with CW driving. 

Figure~\ref{fig4}(b) shows results from more extensive pulse-driving simulations that consider a range of peak driving power values $P_\mathrm{p}$. For each value of $P_\mathrm{p}$, we perform simulations for a range of desynchronisations $\Delta t$ to extract the upper and lower limits of soliton existence. The results are shown as the blue solid curves in Fig.~\ref{fig4}(b). Also shown as orange solid circles are the corresponding upper and lower desynchronisation limits predicted from our CW-driving analysis [specifically using the curve in Fig.~\ref{fig2}(d)]. We observe that the predicted upper limit follows closely the value observed in simulations, whilst the lower limit exhibits a similar offset as observed in Fig.~\ref{fig4}(a). It is worth noting that, whilst the upper desynchronisation limit is expected (and observed) to increase with the driving peak power due to a corresponding increase of the peak background power level [$\Delta t_\mathrm{max} = -V(P_\mathrm{B,peak})$], the lower limit is expected to stay constant at $\Delta t_\mathrm{min} = -V(P_\mathrm{RS,min})$. In practice, our pulse-driven simulations show that this lower limit exhibits a slow decrease with increasing driving peak power, which we attribute to the increasing power gradient experienced by the solitons.

In conclusion, we have shown via numerical simulations that ultrashort dissipative Raman solitons can in principle exist even under conditions of CW driving, provided that the system is operated below the threshold of net Raman gain. Our CW-driving simulations show that the soliton characteristics depend on the power level of the system's homogeneous steady-state. We have leveraged these observations to explain how desynchronisation affects Raman soliton existence and characteristics in pulse-driven configurations. Our analysis yields new insights on the dynamics of ultrashort Raman solitons and explains why steady-state solitons can only be achieved for a finite range of desynchronisations. 

\begin{backmatter}
\bmsection{Funding} We acknowledge financial support from the Marsden Fund of
the Royal Society Te Apārangi of New Zealand.

\bmsection{Data availability} 
 Data underlying the results presented in this paper are not publicly available at this time but may be obtained from the authors upon reasonable request.

\smallskip

\bmsection{Disclosures} The authors declare no conflicts of interest.
\end{backmatter}

\bibliography{sample}

\end{document}